\documentclass[final,3p,times,twocolumn]{elsarticle}

\usepackage{lineno}
\usepackage{amsmath}
\usepackage{hyperref}
\usepackage{extarrows}
\usepackage{amsfonts}
\usepackage{amssymb}
\usepackage{csquotes}
\usepackage{booktabs}
\usepackage{multirow}
\usepackage{threeparttable}
\usepackage{tikz}
\usepackage{pgfplots}
\usepackage{pgf}
\usepackage{pdflscape}
\usepackage{subcaption}
\usepackage{pdfpages}
\usepackage{colortbl}
\usepackage{dcolumn}
\usepackage{upgreek}

\definecolor{Lightgray}{RGB}{235,235,235}
\newcommand{\isotope}[2]{\ensuremath{\mathrm{{}^{#1}#2}}}
\newcommand{\decay}[2]{\xLongrightarrow[\mathrm{{#2}}]{\mathrm{{#1}\,MeV}}}
\newcommand{\decayb}[1]{\xlongrightarrow[\mathrm{{#1}}]{\upbeta}}

\journal{Journal of Applied Radiation and Isotopes}

\begin{document}

\begin{frontmatter}

\title{Production and characterization of a $^{222}$Rn-emanating stainless steel source}

\author[a]{Florian J\"org\corref{corrauth}}
\cortext[corrauth]{Corresponding author}
\ead{fjoerg@mpi-hd.mpg.de}
\author[a,b]{Guillaume Eurin}
\author[a]{Hardy Simgen}
\address[a]{Max-Planck-Institut f\"ur Kernphysik, Saupfercheckweg 1, 69117 Heidelberg, Germany}
\address[b]{\emph{Present Address:} IRFU, CEA, Universit\'e Paris-Saclay, F-91191 Gif-sur-Yvette, France}

\begin{abstract}
Precise radon measurements are a requirement for various applications, ranging from radiation protection over environmental studies to material screening campaigns for rare-event searches.
All of them ultimately depend on the availability of calibration sources with a known and stable radon emanation rate.
A new approach to produce clean and dry radon sources by implantation of $^{226}$Ra ions into stainless steel has been investigated.
In a proof of principle study, two stainless steel plates have been implanted in collaboration with the ISOLDE facility located at CERN.
We present results from a complete characterization of the sources. 
Each sample provides a radon emanation rate of about 2 Bq, which has been measured using electrostatic radon monitors as well as miniaturized proportional counters. 
Additional measurements using HPGe and alpha spectrometry as well as measurements of the radon emanation rate at low temperatures were carried out. 
\end{abstract}


\end{frontmatter}


\vspace*{11.6cm}
\hspace{0.5\textwidth}
\begin{minipage}{0.325\textwidth}
\small{This manuscript version is made available under the CC-BY-NC-ND 4.0 license.}
\footnotesize{\href{https://creativecommons.org/licenses/by-nc-nd/4.0/}{creativecommons.org/licenses/by-nc-nd/4.0/}}
\end{minipage}%
\begin{minipage}{0.15\textwidth}
\hspace*{0.2cm}
\includegraphics[width=0.9\textwidth]{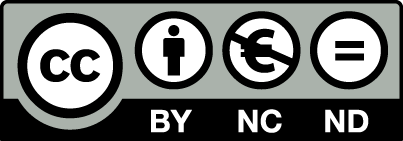}\newline
\end{minipage}%
\vspace*{-11.6cm}
\vspace{-5\baselineskip}


\section{Introduction}
\label{sec:introduction}

The radioactive noble gas radon is a crucial element in various fields of radioactivity measurements. 
In particular its longest-lived isotope $^{222}$Rn significantly contributes to the public's annual  radiation exposure\,\cite{UNSCEAR:2018} and is used as tracer in environmental studies\,\cite{Baskaran:2016}. 
Moreover, it is a severe source of background in fundamental physics experiments searching for rare events at low energies, such as direct dark matter interactions\,\cite{Undagoitia:2015gya,Schumann:2019eaa}.
One example of such experiments are liquid xenon-based detectors\,\cite{Aprile2017,Akerib2020,Zhang:2018xdp,XENON:2020kmp}, which rely on extensive material screening campaigns to select the radio-purest construction materials~\cite{Aprile:2020rn, LZ:2020fty, Aprile:2021xx,PANDAX:2021qiu}.
Moreover, novel radon mitigation methods are being developed to suppress residual radon traces.

All these applications require sensitive and reliable radon detection methods, which in turn need reliable radon emanation sources. 
In this work, we present a novel approach to produce such a source by implantation of $^{226}$Ra into stainless steel targets.
A similar method has also been applied for the production of $^{222}$Rn-emanating tungsten and aluminum samples in an independent study reported recently in\,\cite{Mertes:2022}.
Besides its application for the characterization and calibration of novel radon detectors, a $^{226}$Ra-implanted source can also be used to study novel radon mitigation techniques using coatings applied to stainless steel surfaces\,\cite{Jorg:2022spz}.
Especially the latter case, requires a source with an unaltered steel surface, in order to not bias the deposition process.
This excludes for example the use of a radon source that is produced via electro-deposition (see e.g.\,\cite{Mertens:2020}).


The article is structured as follows: The implantation of the samples is described in section\,\ref{sec:production}, followed by a measurement of the total amount of implanted activity using alpha and gamma spectrometry (section~\ref{sec:spectrometry}).
Their $^{222}$Rn emanation rate at room temperature as well as at low temperatures is reported in section\,\ref{sec:rn_emanation}.
Shortly after the implantation, contributions from various short lived isotopes have been observed, which are detailed in section~\ref{sec:co_implantation}.

\section{Implantation of the sample at ISOLDE}
\label{sec:production}

\begin{figure}[h]
	\centering
	\includegraphics[width=0.49\textwidth]{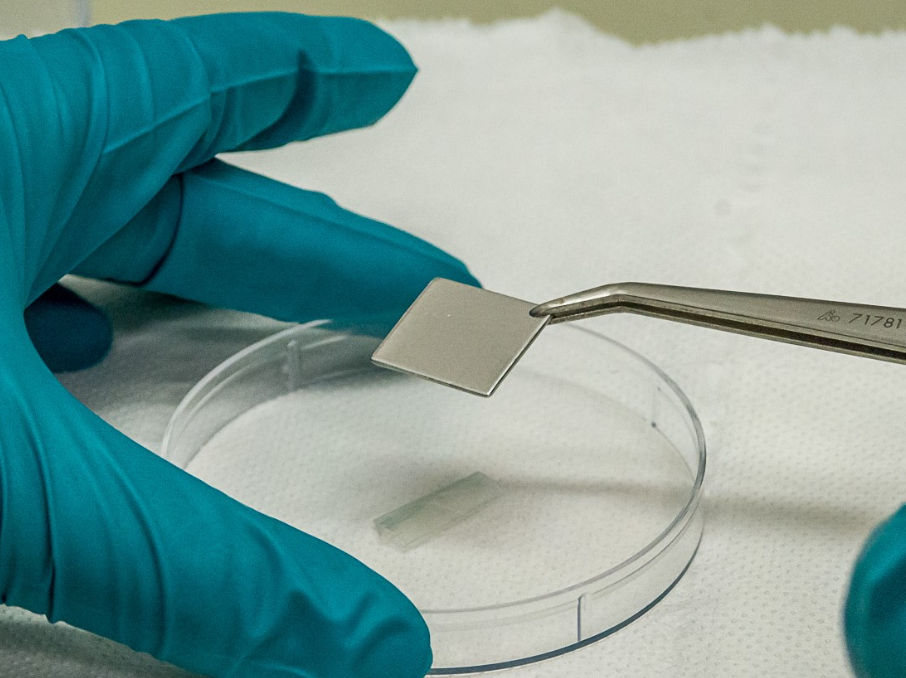}
	
	\hspace{2cm}
	
	\includegraphics[width=0.49\textwidth]{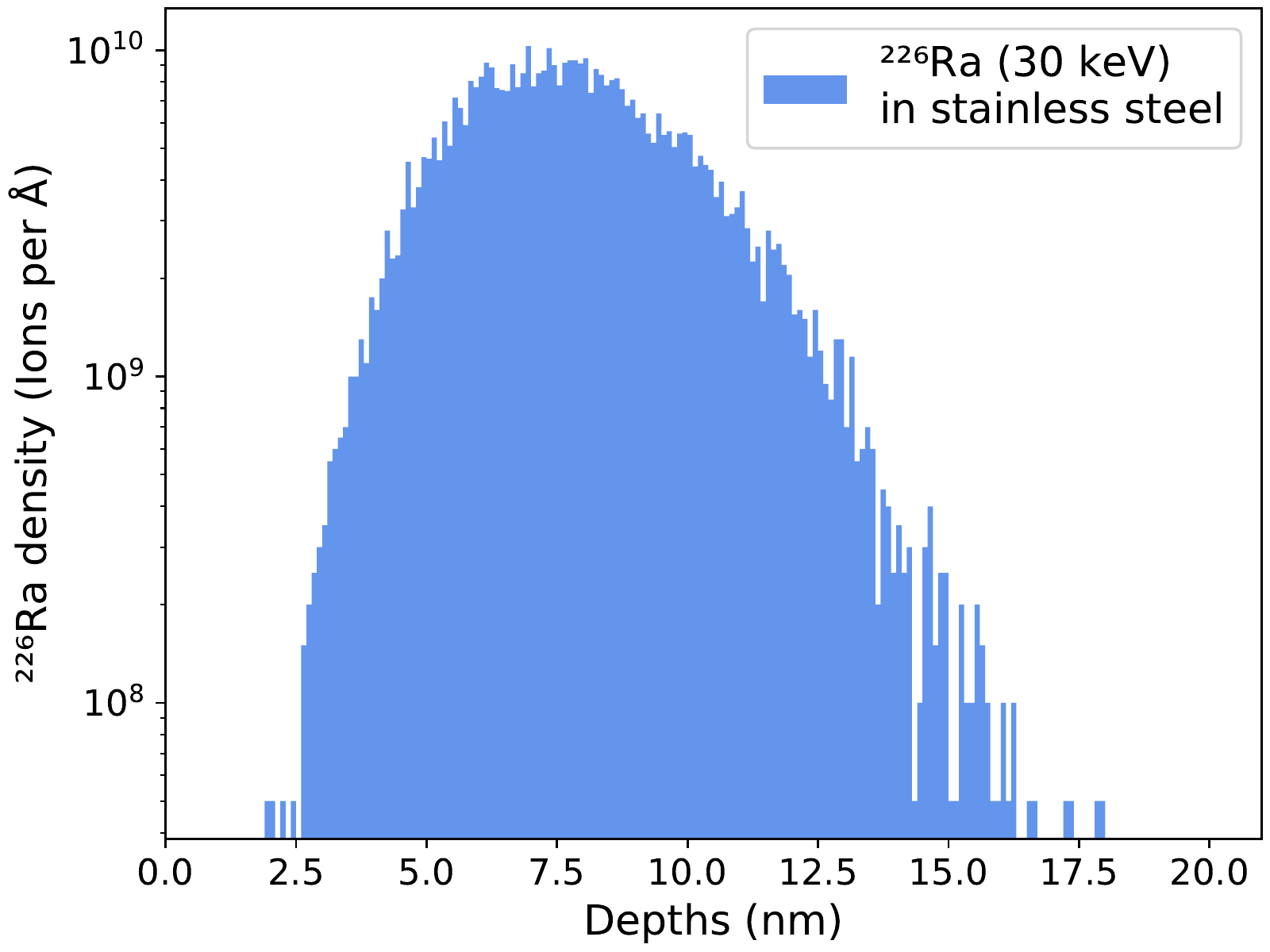}
	\caption{\textbf{Top: }Picture of one of the \isotope{226}{Ra} implanted stainless steel samples. \textbf{Bottom: }Expected implantation depth distribution as simulated using SRIM \cite{Ziegler:2010, Joerg:2017}.}
	\label{fig:isolde_sample_picture}
\end{figure}

The ion separation online device (ISOLDE) is one of the leading infrastructures for the production of radioactive ion beams (RIBs)\,\cite{Kugler:1991tq}.
In a proof of principle study two stainless steel samples (\textit{A} \& \textit{B}) have been implanted with \isotope{226}{Ra}.
One of the 2\,cm by 2\,cm large samples is depicted in the top panel of Figure\,\ref{fig:isolde_sample_picture}.
They were cut  from a 1\,mm thick stainless steel sheet and their surface has been thoroughly de-greased using acetone and 2-propanol prior their implantation.
Each sample got implanted with about $5\times 10^{11}$ radium ions~\cite{Joerg:2017}. 
This translates to an implanted activity of the order of 7\,Bq per sample.
The current of \isotope{226}{Ra} ions achieved throughout the implantation was approximately 3\,pA.
The ions were released from a heated uranium carbide target, which has previously been irradiated by about $1.2\times 10^{18}$ protons.
Given the long half life of \isotope{226}{Ra}, it was possible to perform the implantation off-line i.e. without additional proton irradiation of the target.
The ions were accelerated to a mean energy of 30\,keV and mass separated using ISOLDE's general purpose separator (GPS).
During the implantation process, the ion beam was swept over the samples surface, to cover an area of about 1\,cm by 1\,cm.
The implantation depth profile as simulated using the \enquote{stopping and range of ions in matter} (SRIM) code\,\cite{Ziegler:2010} is shown in Figure~\ref{fig:isolde_sample_picture} (bottom).
From this simulation, it is expected that the radium ions are distributed around an average depth of 7.9\,nm below the surface, with a spread of 2.3\,nm~\cite{Joerg:2017}.

The shallow implantation profile allows for a sufficient radon release, while at the same time it is deep enough to provide sufficient mechanical stability.
The latter was confirmed by a wiping test of \textit{sample A} using an ethanol soaked filter paper.
A subsequent measurement of the radon emanation rate from the wipe revealed a removal of less than 0.2\% of the total activity implanted in the sample.

\section{Spectrometric measurements}\label{sec:spectrometry}

The decay of $^{226}$Ra inside the sample continuously populates the following part of the primordial $^{238}$U decay series:
\begin{align}
\dots~&\isotope{226}{Ra} \decay{4.9}{1600\,y} \isotope{222}{Rn} \decay{5.6}{3.8\,d} \isotope{218}{Po} \decay{6.1}{3.1\,min} \isotope{214}{Pb}\notag\\
&\decayb{27\,min} \isotope{214}{Bi} \decayb{20\,min} \isotope{214}{Po} \decay{7.8}{160\,\upmu s} \isotope{210}{Pb}~\dots\label{eq:rn222_chain}
\end{align}
The alpha transitions of the decay chain are indicated by double arrows and the values of the half-lives and decay energies are taken from\,\cite{Chu:1999}.
Due to the 22\,year long half-life of $^{210}$Pb, the subsequent decays can be neglected in the present case.

The amount of implanted $^{226}$Ra in each sample was measured using alpha and gamma spectrometry.
This allows to disentangle the decays of $^{226}$Ra from the decays of the other isotopes present in the decay chain.
Decays of $^{226}$Ra can be clearly identified via the emitted 4.9\,MeV alpha particle, as well as by the emission of a 186.2\,keV gamma in about 6\% of the decays\,\cite{TabRad_v8}.
For gamma spectrometry, a high-purity germanium (HPGe) detector was used.
Alpha energy spectra have been acquired using different detectors equipped with silicon PIN diodes.
In both measurements, the samples have been placed into a cylindrical high-density polyethylene (HDPE) holder, having an opening ensuring that alpha particles are not stopped after their emission.

\subsection{Alpha spectrometry}\label{subsec:alpha_spectrometry}

The alpha spectrometers used in this work use a windowless Hamamatsu S3204-09 Si-PIN diode with a sensitive area of $\mathrm{18\times 18\,mm^2}$\,\cite{hamamatsu:2021} to measure the energy of impinging alpha particles.
The diode is mounted on the top flange of an evacuated vessel and faces the implanted side of the sample.
To prevent the energy loss of the alpha particles on their way from the sample to the diode, the pressure in the spectrometer is kept below the millibar level.
The spectrometer's relative energy resolution amounts to about 2\%.
It is dominated by the energy loss of the alpha particles in the approximately 100\,nm thick dead layer of the Si-PIN diode\,\cite{Spencer:2004}.
The additional energy loss due to the approximately 8\,nm thin stainless steel layer, covering the implantation site, amounts to only 0.1\%.
Since this is much smaller than the spectrometer's energy resolution, this cannot be applied for the reconstruction of the depth distribution of the implanted $^{226}$Ra ions in the sample.

\begin{figure}[h]
	\centering
    \includegraphics[width=0.49\textwidth]{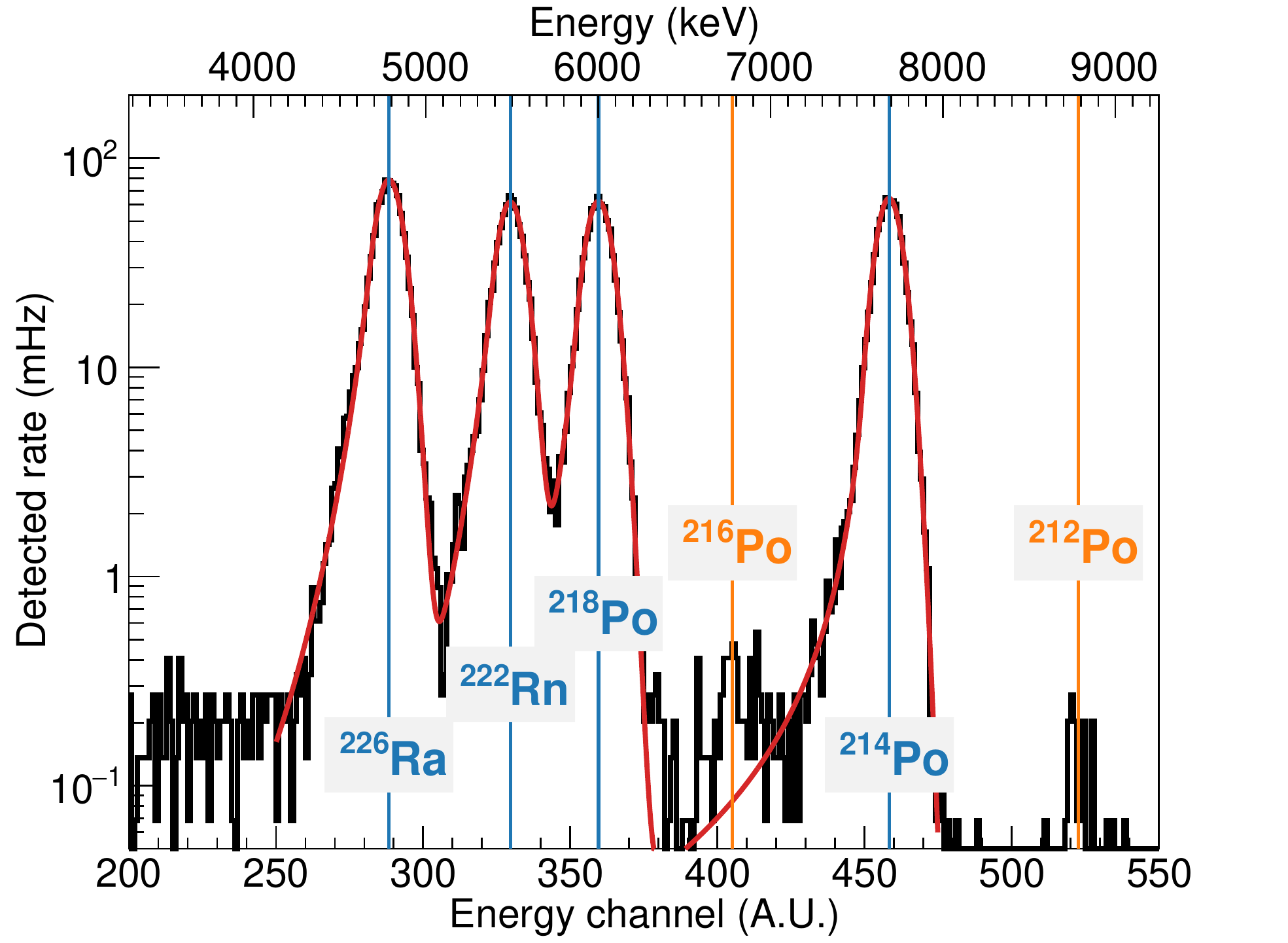}	
	\caption{Alpha spectrum acquired for \textit{sample A} long (1.7\,years) after the implantation~\cite{Lecher:2019}. Alpha emission lines from isotopes belonging to the uranium (blue) and thorium (orange) decay chains are visible.}
	\label{fig:alpha_spectra}
\end{figure}

The diode signal is read-out and amplified before being digitized by a multi-channel analyzer.
An exemplaric alpha spectrum of \textit{sample A} 1.7\,years after the implantation is shown in Figure~\ref{fig:alpha_spectra}.
The alpha emission lines of \isotope{226}{Ra}, \isotope{222}{Rn}, \isotope{218}{Po} and \isotope{214}{Po} are clearly visible.
The additional smaller contributions to the spectrum, which can be attributed to isotopes of the primordial $^{232}$Th decay chain, are discussed further in section~\ref{sec:co_implantation}.
The spectrum is fit with a sum of four individual Crystal Ball functions~\cite{Skwarnicki:1986xj} (red line).

This measurement was performed using the alpha spectrometer described in~\cite{Herrero:2018, Bruenner:2020} and is originally reported in\,\cite{Lecher:2019}.
The sample holder was placed onto the tray of the spectrometer, such that the distance between the sample and the silicon PIN diode amounted to $\mathrm{(1.2 \pm 0.2)\,cm}$.
The geometric detection efficiency for this measurement is estimated using a Monte-Carlo simulation to about $\mathrm{\left(11.2\,^{+2.9}_{-2.2}\right)\%}$, where the uncertainty is dominated by the uncertainty of the distance measurement.

The detected activity of each isotope is extracted via the normalization constant of the respective Crystal Ball function used to describe its emission line.
After correcting for the geometric detection efficiency, the absolute amount of implanted $^{226}$Ra activity into \textit{sample A} can be determined to be $\mathrm{\left(8.70 \pm 0.06\,(stat)~^{+2.0}_{-1.8}\,(syst)\right)\,Bq}$.
The activities of  $^{222}$Rn, $^{218}$Po and $^{214}$Po are found to be lower by about 2\,Bq.
This is in good agreement with the $^{222}$Rn emanation rate of the samples, as will be described in more detail in section\,\ref{sec:rn_emanation}.

A similar measurement has been carried out with \textit{sample B} using a similar alpha spectrometer. 
The distance between the sample and the diode has been increased to $\mathrm{(10.7 \pm 0.3)\,cm}$ for that measurement, resulting in a much lower detection efficiency of only $\mathrm{\left(2.23\,^{+0.11}_{-0.15}\right)\,\times 10^{-3}}$.
At the same time, the larger distance significantly reduces the uncertainty due to distance variations.
The total amount of implanted \isotope{226}{Ra} activity in \textit{sample B} is found to be $\mathrm{\left(9.13 \pm 0.10\,(stat)~^{+0.7}_{-0.4}\,(syst)\right)\,Bq}$.

The activity implanted into both samples is found to be in agreement within their combined uncertainties.
For both samples, it has also been confirmed, that none of the implanted activity extends beyond the $\mathrm{(1.9\pm0.2)\,cm}$ opening diameter of the HDPE holder\,\cite{Lecher:2019}.

\subsection{Gamma spectrometry}\label{subsec:gamma_spectrometry}
\begin{figure}[h]
\centering
    \includegraphics[width=0.35\textwidth]{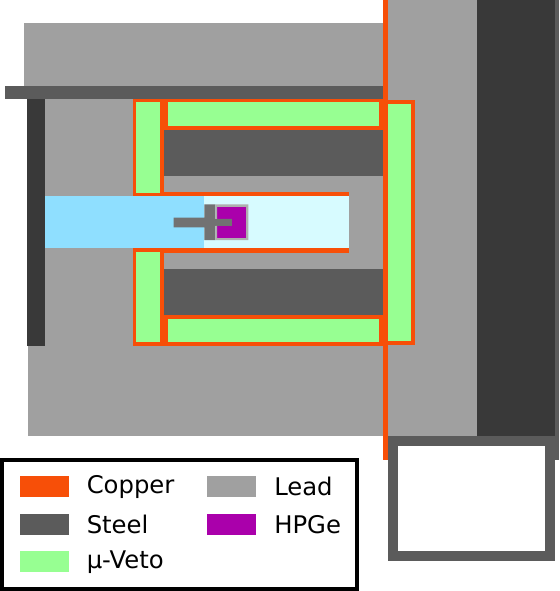}%
    \caption{Cutaway view of the interleaved shielding of the HPGe spectrometer used in this work (figure adapted from~\cite{budjavs:2009}).}\label{fig:hpge_spectrometer} 
\end{figure}

Besides the alpha spectrometric measurements, both samples have been measured in the HPGe spectrometer described in\,\cite{budjavs:2009}.
It has a nitrogen-purged, $\mathrm{120\,cm^3}$ large measurement chamber, housing a 0.63\,kg p-type germanium crystal.
The interleaved copper, steel and lead shielding has a minimum thickness of 16\,cm and is sketched in Figure~\ref{fig:hpge_spectrometer}.
The detector is operated in the shallow underground laboratory at the Max-Planck-Institut f\"ur Kernphysik in Heidelberg (MPIK) at a depth of 15 meters water equivalent\,\cite{Laubenstein:2004}.
During the measurement, the sample inside the HDPE holder is placed directly in front of the crystal.
The combined geometric and full-absorption probability for 186.2\,keV gamma rays emitted by the sample is determined using a GEANT4-based\,\cite{Agostinelli:2002hh} simulation to be $(10.2\pm1.2)$\% as detailed in\,\cite{budjavs:2009}.

\begin{figure}[h]
	\centering
	\includegraphics[width=0.45\textwidth]{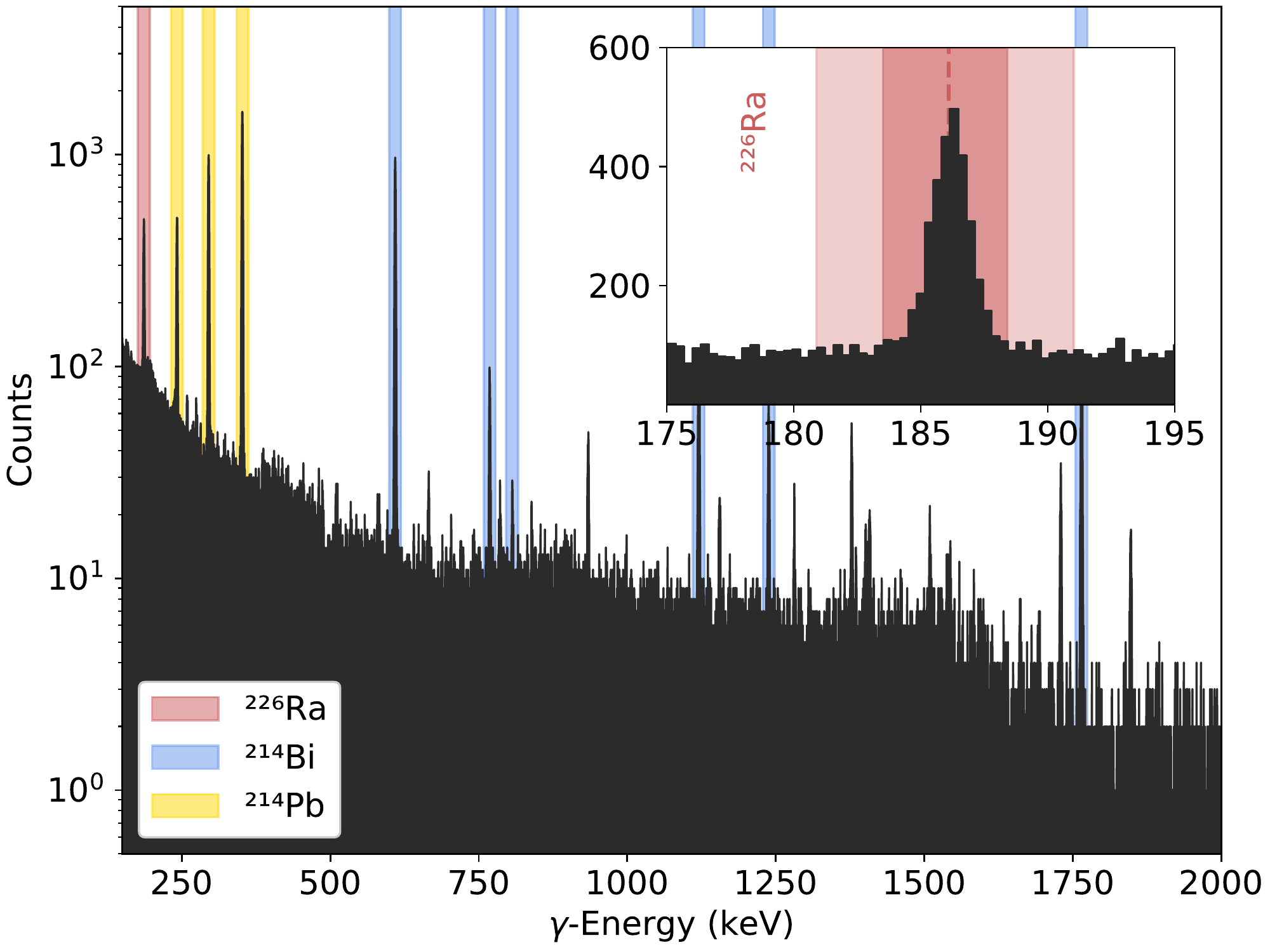}
	\caption{HPGe gamma spectrum acquired for \textit{sample B}. The most prominent emission lines of $^{226}$Ra (red) as well as its short-lived daughter isotopes $^{214}$Bi (blue) and $^{214}$Pb (yellow) are highlighted. The region around the 186.2\,keV emission line of $^{226}$Ra is shown enlarged in the inset.}
	\label{fig:gamma_spectrum}
\end{figure}

Figure\,\ref{fig:gamma_spectrum} shows the gamma spectrum obtained from \textit{sample B} 3.4\,years after the implantation.
Shown in the inset is a zoomed-in region around the 186.2\,keV gamma emission line.
The activity is determined from the full-absorption region (darker red), while the line background is estimated and subtracted using the side bands (light red).
Implanted $^{226}$Ra activities of $\mathrm{\left(7.4\pm 0.1\,(stat)\pm 0.9\,(syst)\right)\,Bq}$ and $\mathrm{\left(8.4\pm 0.3\,(stat)\pm 1.0\,(syst)\right)\,Bq}$ are found for \textit{sample A} and \textit{sample B} respectively.

The spectrum also shows the emission lines of the \isotope{222}{Rn} daughter isotopes $^{214}$Bi and $^{214}$Pb.
Similar to the alpha spectrometric measurement, their activities are found to be lower by about 2\,Bq, since the emanated radon is purged from the measurement chamber by the flow of clean nitrogen.
The exact quantification of the radon emanation rate will be described in the following section.

\section{Radon emanation measurements}\label{sec:rn_emanation}

\begin{figure}[h]
	\centering
	\includegraphics[width=0.4\textwidth]{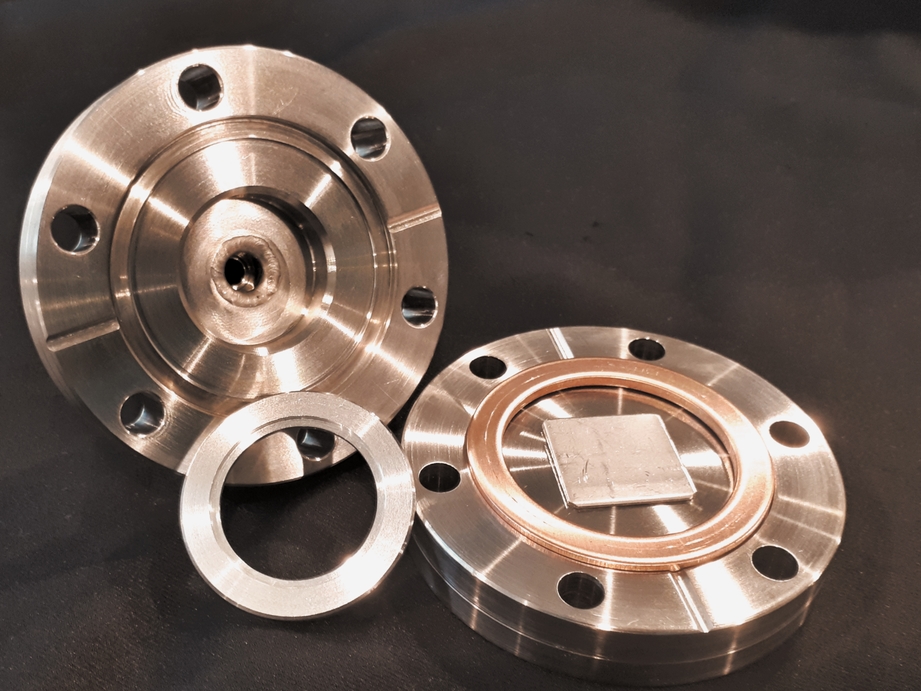}
	\caption{Emanation vessel used for the measurement of the sample's radon emanation rates. The aluminum ring (front) is used to fix the sample in place.}
	\label{fig:thermal_holder}
\end{figure}

The radon release from the samples has been the main objective for their production.
For its measurement, the sample is enclosed in the emanation vessel depicted in Figure\,\ref{fig:thermal_holder}.
It is made from two CF-40 vacuum flanges, between which the sample is held in place using an aluminum ring.
The opening diameter of the ring is $\mathrm{24.3\,mm}$, such that it does not cover the implanted surface.
For the emanation, the vessel is filled with helium at a pressure of at least 200\,mbar.
The top flange (Figure\,\ref{fig:thermal_holder}, left) features a concentric recess, such that a minimum clearance of about 5.6\,mm between the sample surface and the flange material is guaranteed.
It has been verified using the SRIM code\,\cite{Ziegler:2010}, that this distance is sufficient to prevent the recoil implantation of $^{222}$Rn nuclei into the container walls for helium filling pressures of at least 100\,mbar.
This is important, since those implanted nuclei would likely be lost for the subsequent radon measurement.

After an emanation time of several days, the accumulated radon is extracted together with the helium carrier gas.
The radon is then collected on an activated carbon trap, which is cooled down to liquid nitrogen temperature.
Following the end of the extraction, the trap is heated up and the radon is transferred into a radon detector for its measurement.
Afterwards, the detected activity is corrected to account for the incomplete accumulation and decay, as well as for the decay of radon during the transport of the sample.

\subsection{Room temperature emanation rate}\label{subsec:rt_rn_emanation}

\begin{figure}[h]
	\centering
	\includegraphics[width=0.49\textwidth]{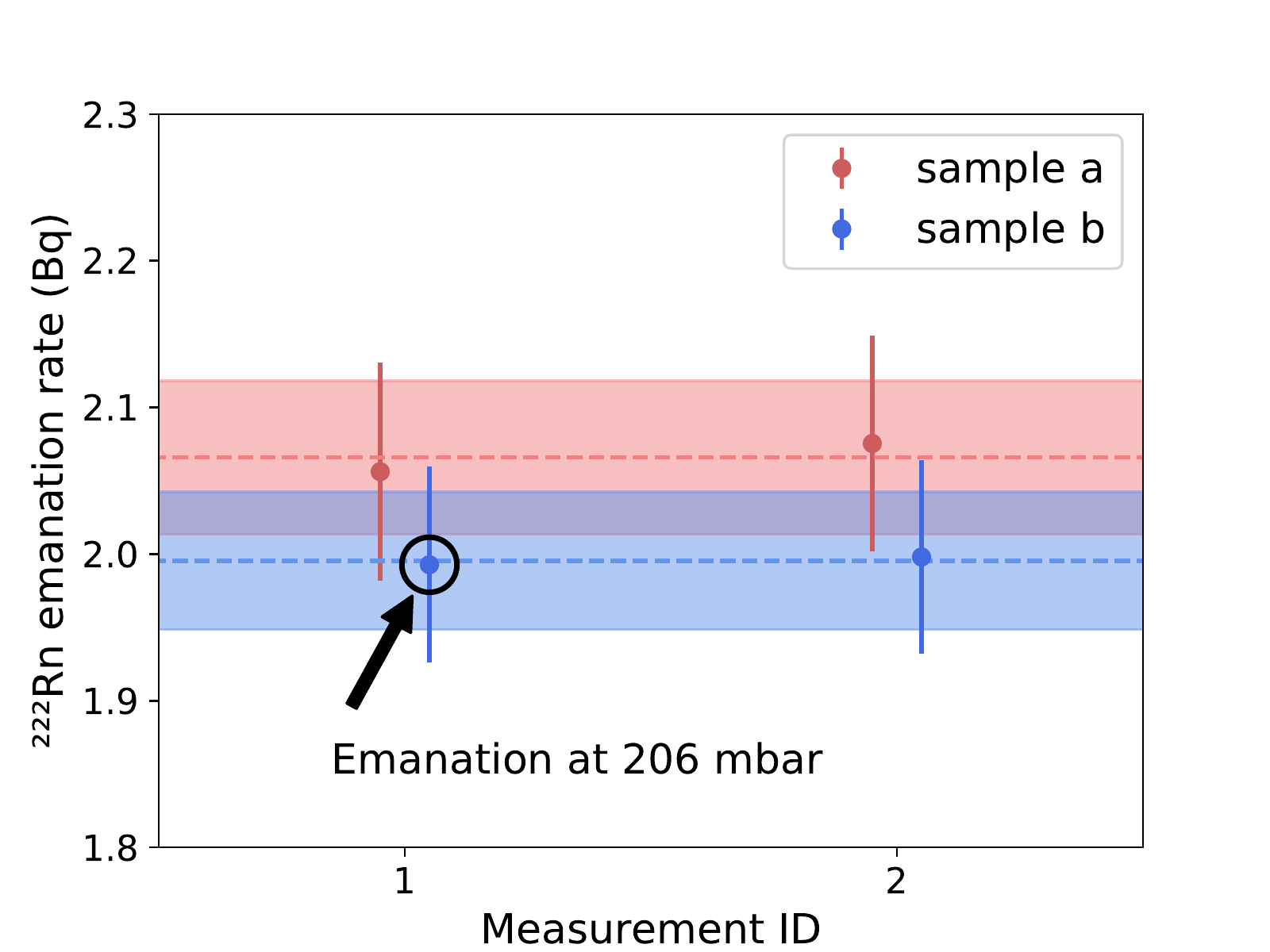}
	\caption{Equilibrium emanation rate of \isotope{222}{Rn} at room temperature for both samples. Measurements carried out in helium atmosphere at pressure of 1050\,mbar (except for indicated measurement) using miniaturized proportional counters~\cite{Kiko:2001}.}
	\label{fig:room_temp_emanation}
\end{figure}

The room temperature emanation rate has been measured using miniaturized proportional counters as described in \cite{Aprile:2020rn, Zuzel:2009}.
Each sample has been measured twice, where both results are found to be well in agreement, as illustrated in Figure\,\ref{fig:room_temp_emanation}.
Three out of the four measurements were carried out with a helium pressure of 1050\,mbar in the emanation vessel.
For the first measurement of \textit{sample B}, a much lower pressure of only 200\,mbar has been chosen.
Since this measurement yielded the same result, a strong pressure dependence of the radon emanation rate can be excluded.

The mean radon emanation rates of both samples are found to be to $\mathrm{(2.07\pm0.05)\,Bq}$ and $\mathrm{(2.00\pm0.05)\,Bq}$ for \textit{sample A} and \textit{sample B} respectively.
This represents 20\% to 28\% of the rate at which $^{222}$Rn is produced inside the sample (see previous section).
This emanation fraction is found to be well explained by a purely recoil-driven emanation process.
From the mean implantation depth of 7.9\,nm, as well as the mean range of the $^{222}$Rn recoil in stainless steel (14.8\,nm), an emanation fraction of 21.5\% would be expected.
More details on this estimation are provided in\,\cite{Jorg:2022spz}.

\subsection{Emanation rate at low temperatures}\label{subsec:cryo_rn_emanation}

\begin{figure}[h]
	\centering
	\includegraphics[width=0.4\textwidth]{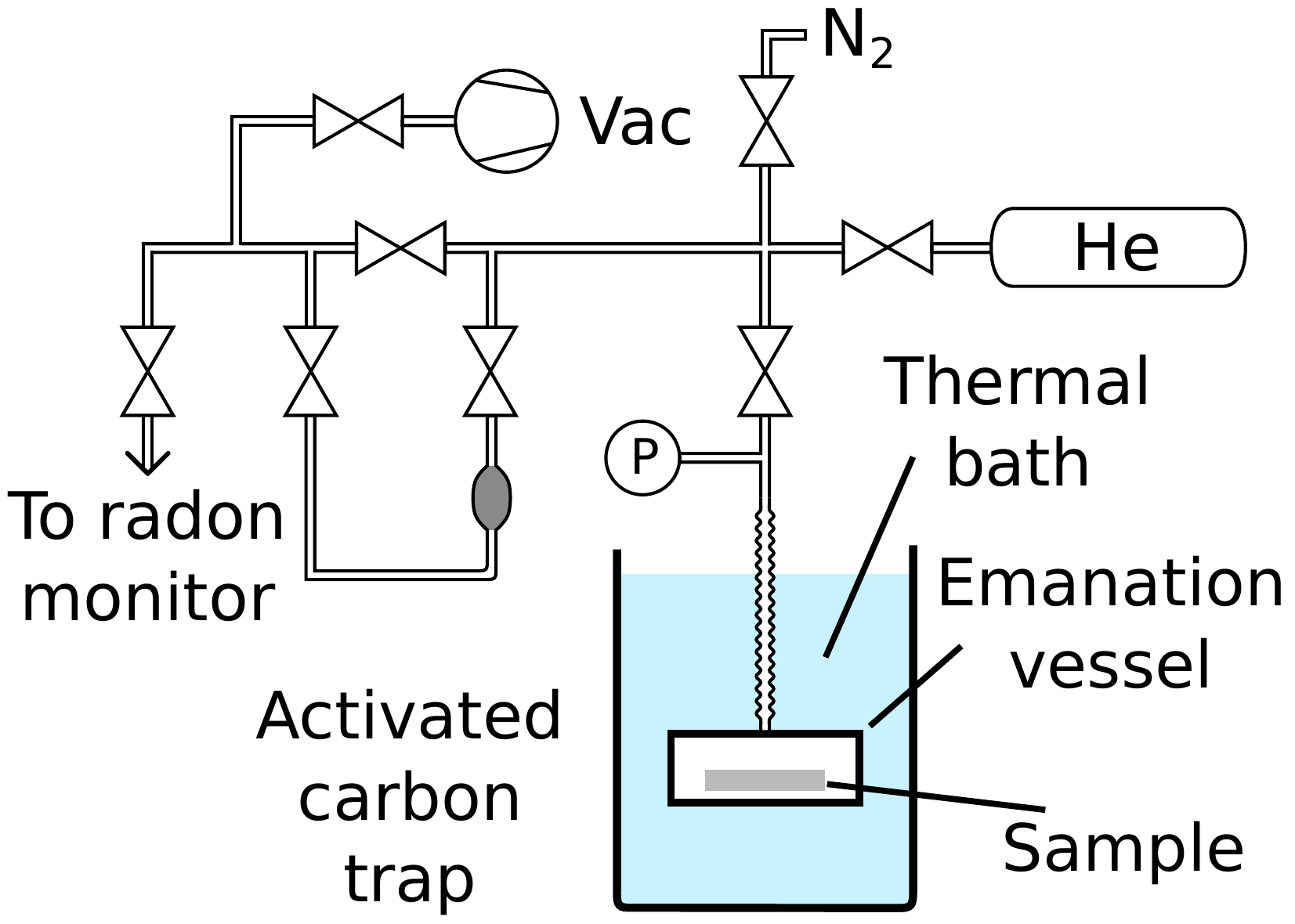}%
	\caption{Sketch of the setup used for the study of temperature dependence of radon emanation rate.}
	\label{fig:thermal_setup}
\end{figure}

The dominant emanation mechanism can also be determined by investigating the temperature dependence of the radon emanation rate.
While a diffusion driven emanation process would exhibit a strong temperature dependence, its absence would point towards a recoil dominated emanation.
Therefore, and for possible applications of the sources in a cryogenic environment, the temperature dependence of the radon emanation rate from \textit{sample B} has been measured using the setup sketched in Figure\,\ref{fig:thermal_setup}.
The emanation vessel is submersed in a thermal bath with 4\,liters of a 2:1 mixture of ethylene glycol and water, allowing to reach temperatures as low as $\mathrm{-30\,^\circ C}$.
The bath temperature is maintained by a HAAKE EK 90 immersion chiller and homogenized by a magnetic stirrer.
A 30\,cm long bellow connects the emanation vessel to the rest of the setup to reduce the external heat input.
For each measurement, the emanation vessel was filled with helium to a pressure of $(203\pm3)\,\mathrm{mbar}$.
After several days, the radon is extracted via an activated carbon trap and transferred into the electrostatic radon monitor described in\,\cite{Kiko:2001,Bruenner:2017}.

It consists of a 4\,liter large hemispheric vessel, equipped with the same windowless Si-PIN diode as is used for the alpha spectrometer (see section\,\ref{subsec:alpha_spectrometry}).
During the measurement, the detector is filled with nitrogen to a pressure of 1050\,mbar and the diode is biased with a negative high-voltage of 1\,kV. 
The resulting electric field points from the grounded vessel towards the diode, allowing to collect the predominantly positively charged radon daughters\,\cite{Pagelkopf2003}. 
Their subsequent alpha decays can then be registered by the diode.
The combined collection and detection efficiency has been determined to be $\mathrm{(35 \pm 2)\,\%}$ using a reference source with a radon emanation rate of $\mathrm{(52 \pm 3)\,mBq}$, which has been determined using the miniaturized proportional counters introduced above.

\begin{figure}[h]
	\centering
	\includegraphics[width=0.49\textwidth]{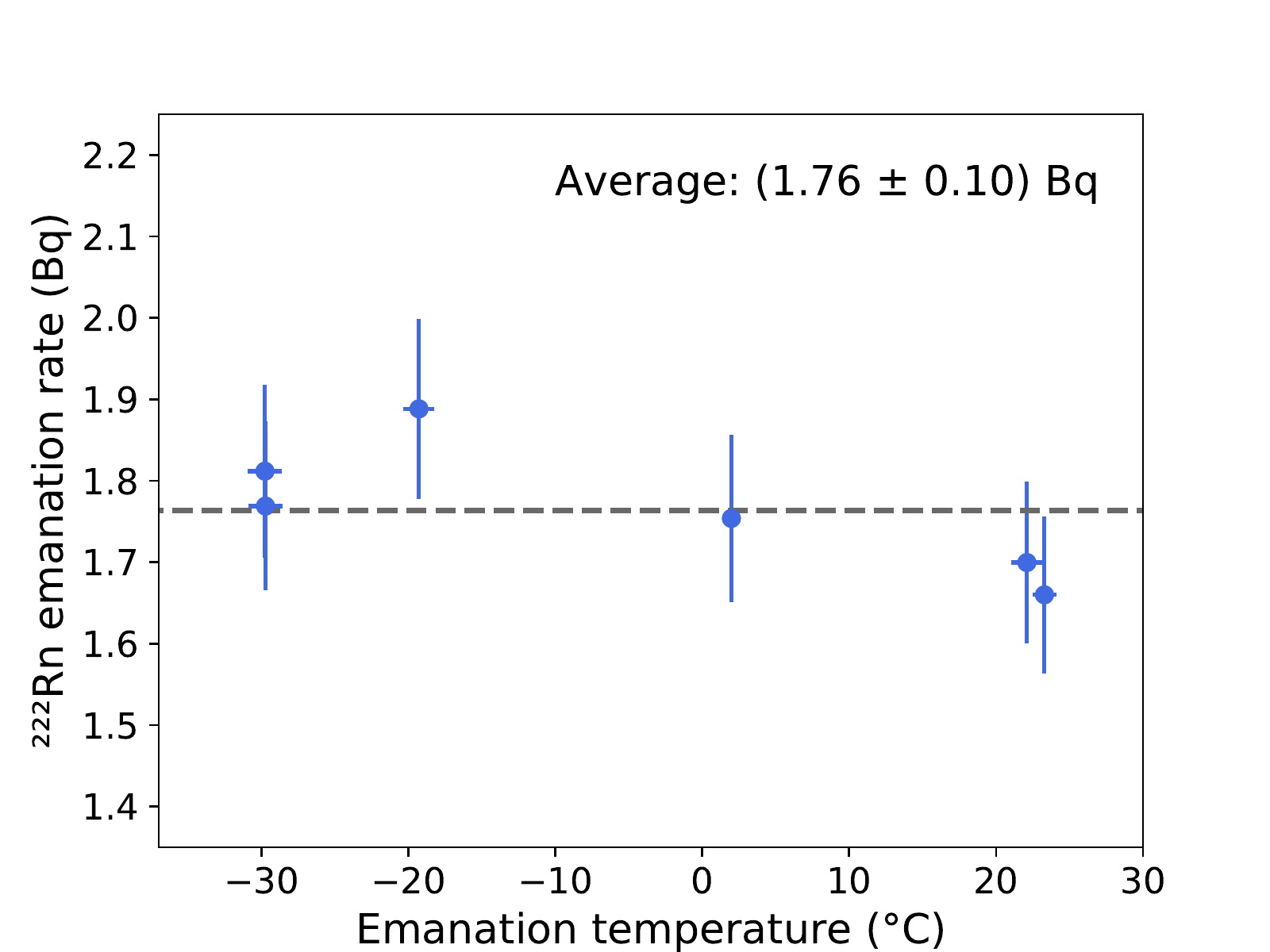}%
	\caption{Temperature dependence of the radon emanation rate of the \isotope{226}{Ra} implanted stainless steel \textit{sample B} as measured using an electrostatic radon monitor.}
	\label{fig:thermal_result}
\end{figure}

Figure~\ref{fig:thermal_result} shows the absolute radon emanation rate of the implanted \textit{sample B} at different emanation temperatures.
Here the measurements at room temperature used the same setup as described above, but with the thermal bath removed.
The measurement uncertainty contains the statistical uncertainty from the counting of alpha events, time variations in the extraction and filling procedure as well as the uncertainty associated to the determination of the detection efficiency of the radon monitor.
Given the overall uncertainty of the measurement, the \isotope{222}{Rn} emanation rate of the sample is found to be stable with respect to the emanation temperature at a 5\% level for temperatures above $\mathrm{-30\,^\circ C}$.

The average radon emanation rates measured by the proportional counters and the electrostatic radon monitor are found to differ slightly, while the latter are lower by about 12\%.
Possible reasons for this include a loss of $^{222}$Rn during the transfer procedure as well as a slight overestimation of the collection and detection efficiency of the monitor.
Furthermore, the elevated radon activity during the measurement could lead to charge accumulation, which in turn diminishes the electrostatic collection efficiency.
Because the rate during the calibration measurement is lower by almost a factor of 50, this effect would not be accounted for.

\section{Observation of short-lived isotopes}\label{sec:co_implantation}

The measurements presented up to now have been carried out more than one year after the implantation has been performed.
The measurements which have been carried out shortly after the implantation revealed the presence of several short lived contaminants on the samples which have been introduced during the implantation process.

\begin{figure}[h]
	\centering
	\includegraphics[width=0.475\textwidth]{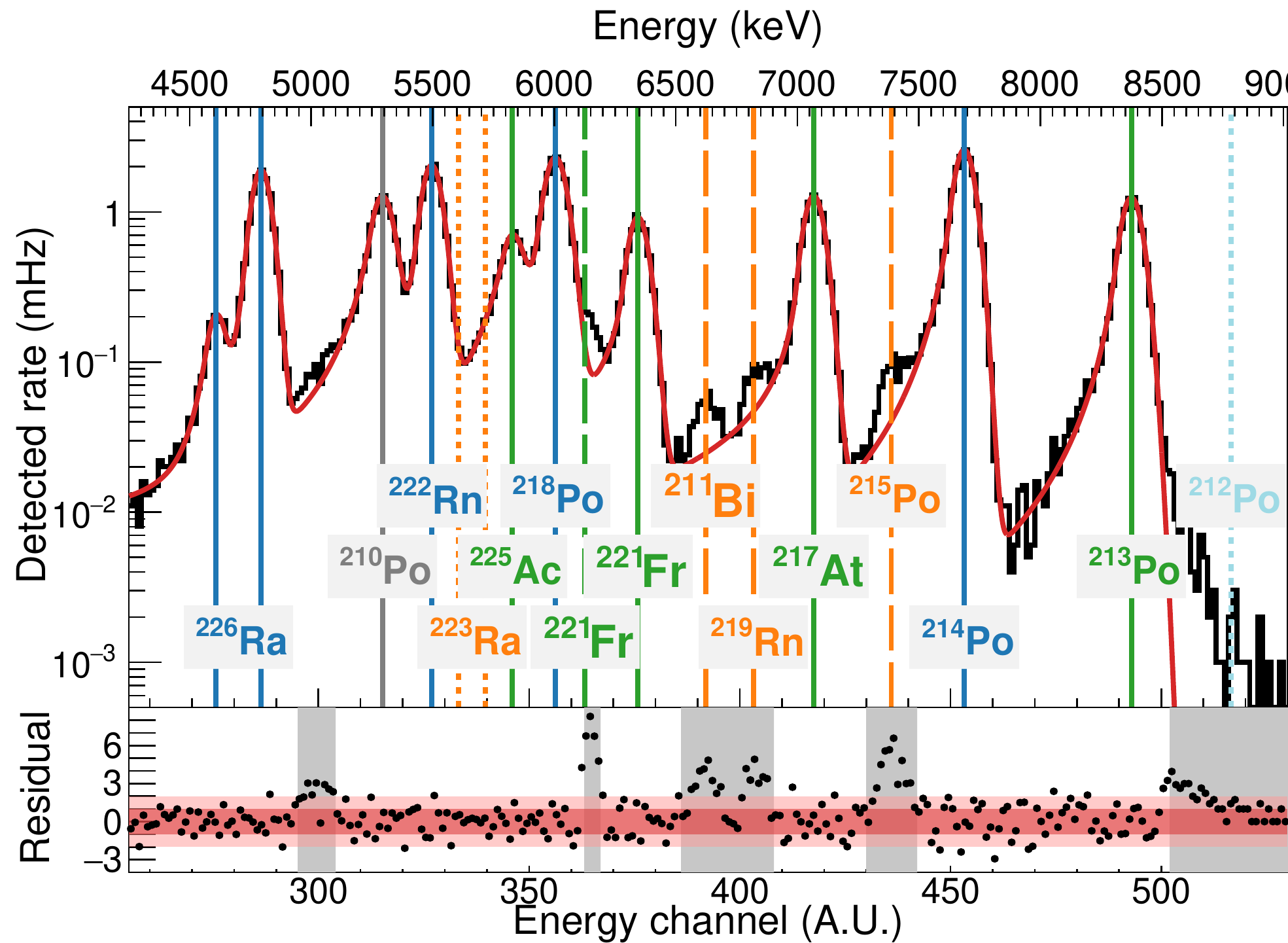}%
	\caption{Alpha emission spectrum of the implanted stainless steel \textit{sample B} 13\,weeks after the implantation~\cite{Herrero:2018}. Most intense alpha emission lines (solid lines) are fit using a sum of Crystal Ball functions~\cite{Skwarnicki:1986xj} with the residuals between data and fit function shown in the lower panel. Data falling in the gray bands were excluded from the fit. Dashed lines indicate possible subdominant contributions.}
	\label{fig:isolde_alpha_spectrum}
\end{figure}

Figure\,\ref{fig:isolde_alpha_spectrum} shows an alpha spectrum of \textit{sample B} acquired only 13 weeks after the implantation\,\cite{Herrero:2018}.
For this measurement the sample was placed at a distance of $\mathrm{(13.2\pm 0.2)\,cm}$ with respect to the Si-PIN diode and the pressure inside the spectrometer was kept at a sub millibar level by continuous pumping.
The complete spectrum is fit using a sum of 10 individual Crystal Ball functions~\cite{Skwarnicki:1986xj} modeling each peak separately.
The distribution of residuals shown in the lower panel of Figure~\ref{fig:isolde_alpha_spectrum} indicates reasonable agreement between the fit and the data, except within the regions indicated by the gray bands.
Those are excluded from the fit in order to reduce bias from the subdominant peaks in the spectrum.
Contributions from at least two radioactive decay chains can be clearly identified by the alpha emission lines of their respective daughter nuclei. Their presence has also been confirmed by independent gamma spectrometry measurements.
Isotopes belonging to the decay chain of \isotope{226}{Ra} (see section\,\ref{sec:spectrometry},  equation\,\ref{eq:rn222_chain}) are highlighted by solid blue lines in Figure~\ref{fig:isolde_alpha_spectrum}.

Whereas solid green lines indicate decay products belonging to the decay chain of \isotope{225}{Ra}.
\begin{align*}
\isotope{225}{Ra} &\decayb{15\,d} \isotope{225}{Ac} \decay{5.9}{10\,d} \isotope{221}{Fr} \decay{6.5}{4.9\,min} \isotope{217}{At}\\
&\decay{7.2}{32\,ms} \isotope{213}{Bi} \decayb{46\,min} \isotope{213}{Po} \decay{8.5}{4.2\upmu s} \isotope{209}{Pb}
\end{align*}
The visible contribution from \isotope{210}{Po} (gray line) is a known background contamination coming from previous measurements and is not representative for the sample.
Furthermore, few subdominant peaks are visible in the spectrum.
Three of these peak candidates fit the emission energies of isotopes from the \isotope{227}{Th} decay chain and are indicated using dashed orange lines.

\begin{align*}
    \isotope{227}{Th} &\decay{6.0}{19\,d} \isotope{223}{Ra} \decay{5.7}{11\,d} \isotope{219}{Rn} \decay{6.8}{4.0\,s} \isotope{215}{Po}\\ 
&\decay{7.4}{1.8\,ms} \isotope{211}{Pb} \decayb{36\,min} \isotope{211}{Bi} \decay{6.6}{2.1\,min} \isotope{207}{Tl}
\end{align*}

The alpha emission lines of \isotope{227}{Th} itself as well as \isotope{223}{Ra} cannot be observed, because they are covered by other isotopes with higher count rates. 
A fourth contribution from isotopes belonging to the primordial \isotope{232}{Th} decay chain can be seen by the peaks from $^{212}$Po and $^{216}$Po in the alpha spectrum shown in Figure~\ref{fig:alpha_spectra}.
This is confirmed independently by the gamma emission line of $^{208}$Tl at an energy of 2.61\,MeV.
Since these isotopes are found even long after the implantation, $^{224}$Ra with a half-life of 3.7\,days can be excluded as their origin.
Therefore, either $^{228}$Th ($T_{1/2} = 1.9\,y$) and/or $^{228}$Ra ($T_{1/2} = 5.8\,y$) needs to have been co-implanted into the samples.

Table\,\ref{tab:short_lived} summarizes the inferred activities of short-lived isotopes present in the sample, right after the implantation at an order of magnitude level.
More details on this estimation are provided in\,\cite{Jorg:2022spz}.

\begin{table}[h]
	\centering
	\begin{threeparttable}
		\caption{Overview over the implanted activities found in short-lived isotopes as well as the respective ratio of implanted ions to the number of implanted $^{226}$Ra ions (beam purity).}\label{tab:short_lived}
		\centering
		\begin{tabular}{lcc}
			\toprule
			Isotope              &  $\mathrm{N_{ions}}$       &  Initial activity\\
			\midrule
			\rowcolor{gray!5}
			\isotope{226}{Ra}    &  $10^{12}$	&   $\mathrm{\mathcal{O}(10\,Bq)}$\\
            \rowcolor{gray!15}
            \isotope{225}{Ac}    &   $10^{9}$    	&   $\mathrm{\mathcal{O}(1\,kBq)}$\\
            \rowcolor{gray!5}
            \isotope{227}{Th}    &  $10^{7}$	&   $\mathrm{\mathcal{O}(1\,Bq)}$\\
            \rowcolor{gray!15}
            \isotope{228}{Th}/\isotope{228}{Ra}
                                &   $10^{6}$   &   $\mathrm{\mathcal{O}(10\,mBq)}$\\
			\bottomrule
		\end{tabular}
	\end{threeparttable}
\end{table}
The second column of the table reports for each isotope the absolute number of ions that got implanted into the sample.
As expected, the contamination of the RIB decreased with increasing distance of the isotopes to the selected mass number of 226.

\begin{figure}[h]
	\centering
	\includegraphics[width=0.45\textwidth]{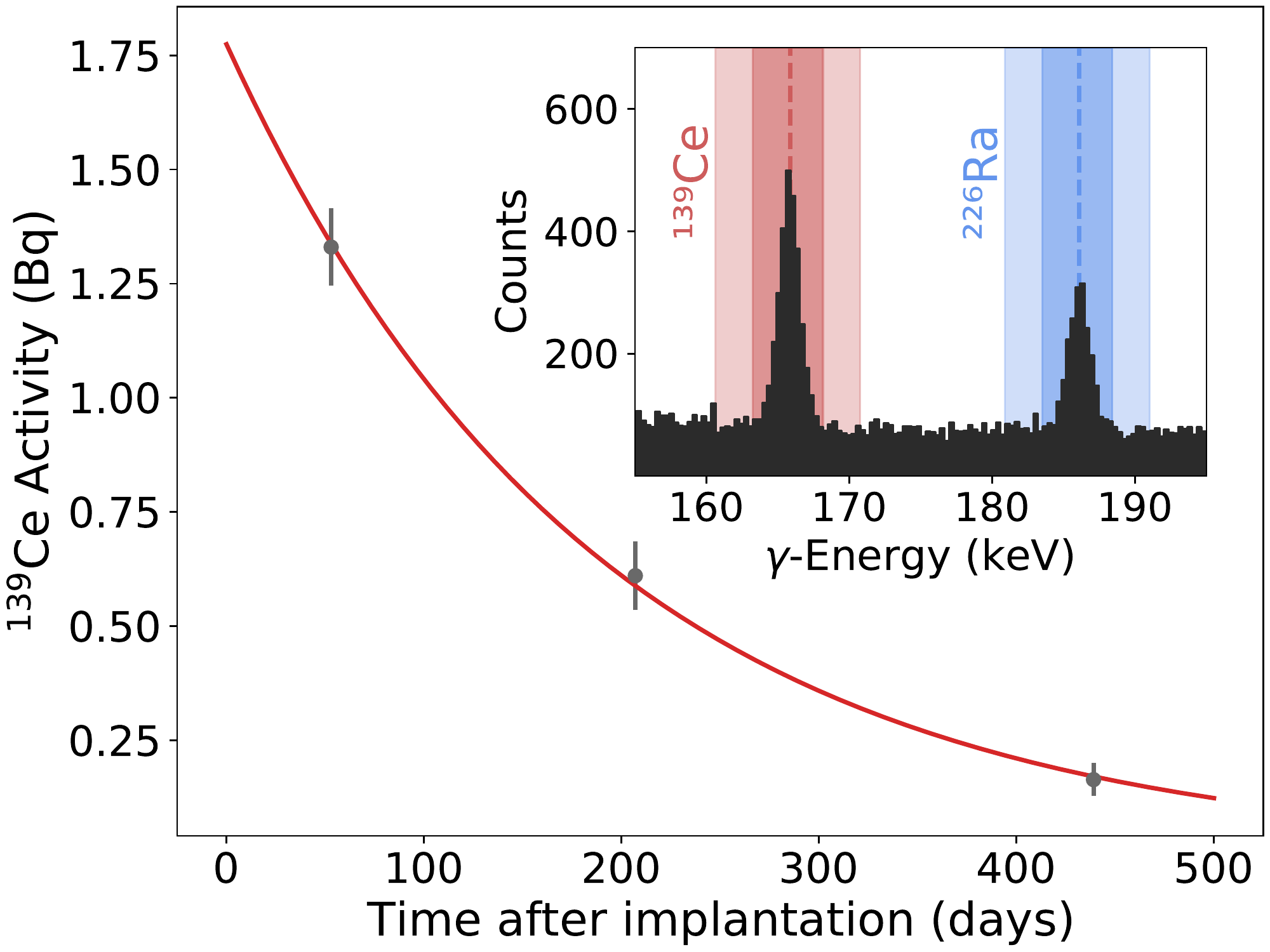}%
	\caption{Time evolution of the $^{139}$Ce activity found in \textit{sample B}. The inset shows the corresponding region of the gamma spectrum for the second measurement.}
	\label{fig:isolde_cerium_evolution}
\end{figure}

The gamma spectra of \textit{sample B} revealed an additional single line at an energy of 165.9\,keV which can be explained by the presence of \isotope{139}{Ce}~\cite{Chu:1999}.
The relevant part of the spectrum is shown in the inset of Figure\,\ref{fig:isolde_cerium_evolution}.
From the decrease of activity found in this line over the course of three measurements over a period of 1 year, a half-life of $\mathrm{(130\pm 5)\,days}$ can be estimated (see red line in Figure\,\ref{fig:isolde_cerium_evolution}). 
This closely fits the literature value of $T_{1/2}(\isotope{139}{Ce}) = 137.6\,\mathrm{days}$~\cite{Chu:1999} which confirms this hypothesis.
Since this isotope far from the selected ion mass, it seems unlikely that it arrived in the sample as an impurity of the RIB.
Furthermore, it is found only in \textit{sample B}, pointing towards a possible contamination during subsequent handling.

However, after a few months of \textit{cool-down} the majority of the remaining activity is from \isotope{226}{Ra} with a trace of activity left from \isotope{228}{Th} and/or \isotope{228}{Ra}.
Therefore, none of these short-lived contaminants had a negative impact on the usability of the samples as sources of radon emanation.

\section{Summary and outlook\label{sec:conclusion}}

In this article a novel method to obtain stainless steel radon sources by implantation of \isotope{226}{Ra} using a radioactive ion beam has been described.
Two such samples have been prepared in collaboration with the ISOLDE facility at CERN.
Both got implanted with a $^{226}$Ra activity of the order of 7\,Bq.
A radon emanation rate of 2\,Bq has been determined for both samples, which is well compatible with the expectation from a recoil-dominated emanation process.
Furthermore, the emanation rate is found to be pressure independent and stable within 5\% for temperatures as low as -30\,$^\circ$C.
Table~\ref{tab:comparison_table} summarizes the key results from different measurement techniques for both samples.

\begin{table}[h]
	\centering
	\begin{threeparttable}
		\caption{Comparison of the results from HPGe spectrometry, alpha spectrometry and radon emanation rates of the two implanted stainless steel samples.}\label{tab:comparison_table}
		\centering
		\begin{tabular}{lll}
			\toprule
			Measurement                     & 	&		Result (Bq)		\\
			\midrule
            \rowcolor{gray!15}
			                                & a     & $2.07\pm 0.03\,\text{\small{(stat)}}\pm 0.04\,\text{\small{(syst)}}$ \\
            \rowcolor{gray!15}
			\multirow{-2}{*}{\isotope{222}{Rn} emanation}
                                            & b     & $2.00\pm 0.03\,\text{\small{(stat)}}\pm 0.04\,\text{\small{(syst)}}$  \\
            \rowcolor{gray!5}
			                                & a     & $7.4\pm 0.1\,\text{\small{(stat)}}\pm 0.9\,\text{\small{(syst)}}$ \\
			\rowcolor{gray!5}
			\multirow{-2}{*}{$\upgamma$-spectrometry}
			                                & b     & $8.4\pm 0.3\,\text{\small{(stat)}}\pm 1.0\,\text{\small{(syst)}}$\\
            \rowcolor{gray!15}
			                                & a     & $8.7 \pm 0.1\,\text{\small{(stat)}}~^{+2.0}_{-1.8}\,\text{\small{(syst)}}$\\
            \rowcolor{gray!15}
			\multirow{-2}{*}{$\upalpha$-spectrometry}
			                                & b     & $9.1 \pm 0.1\,\text{\small{(stat)}}~^{+0.7}_{-0.4}\,\text{\small{(syst)}}$\\
			 \bottomrule
		\end{tabular}
	\end{threeparttable}
\end{table}

Furthermore, an initial activity of several short-lived co-implanted isotopes has been observed in the samples.
However, due to their short half-lives, they did not affect the usability of the samples and were removed by waiting for the sample to cool-down.
Future implantations, however, might make use of a mass filter with higher separation power like the high resolution separator (HRS)~\cite{Kugler:1991tq}, in order to mitigate such contamination.
Additionally, an ionization source with higher chemical selectivity like the resonance ionization laser ion source (RILIS) might be preferred over the surface ionization source used in this study.
The increased radium ionization efficiency of this source\,\cite{Fedosseev:2017zxg}, could reduce the implantation duration.

\section*{Acknowledgements}

We wish to acknowledge the support of the the Max-Planck society as well as the Federal Ministry of Education and Research (BMBF) for equipment used during the implantation.
We want to thank Prof. Dr. Klaus Blaum for being very helpful establishing the contact to the ISOLDE team.
Dr. Karl Johnston, Dr. Juliana Schell, Joao Guilherme Correia as well as Robinson Alves and their colleagues from ISOLDE are acknowledged for carrying out the implantation as well as for fruitful discussions.
Furthermore, we would like to thank Dr. Jochen Schreiner for his support in questions of radio protection, as well as the MPIK technicians for their assistance with the measurements.

\bibliography{manuscript_v1}

\end{document}